\documentclass[
  aps,
  pre,
  reprint,
  amsmath,
  amssymb,
  superscriptaddress,
  floatfix,
  nofootinbib,
  showkeys
]{revtex4-2}

\usepackage{graphicx}
\usepackage{empheq}
\usepackage{xcolor}

\DeclareMathOperator{\sech}{sech}

\begin{document}

\title{On the Kinks in Discrete Systems}

\author{Eugene Kogan}
\email{eugene.kogan@biu.ac.il}
\affiliation{Department of Physics, Bar-Ilan University, Ramat Gan 5290002, Israel}

\date{\today}

\begin{abstract}
We use perturbation theory to study kinks in nonlinear Klein--Gordon
($\phi^4$ and sine-Gordon) chains and in a discrete series-connected
Josephson transmission line. The expansion parameter is the ratio of the
lattice period to the kink width. The next-to-leading-order approximation
modifies the kink profiles obtained previously in the leading-order
approximation.
\end{abstract}

\keywords{discrete systems, kinks, solitons, Josephson transmission line}

\maketitle

\section{Introduction}
\label{introd}

Kinks in discrete systems are topological solitons that interpolate between distinct vacuum states of a lattice, as in discrete $\phi^4$ and sine-Gordon chains. In the $\phi^4$ chain, they describe transitions between the two minima of a double-well potential, whereas in the sine-Gordon chain they correspond to phase shifts of $2\pi$ in a periodic potential. Spatial discreteness breaks the continuous translational symmetry of the corresponding field theory and can therefore modify the kink profile, create a Peierls--Nabarro (PN) potential, and couple a moving kink to linear lattice waves.

The perturbative treatment of soliton motion developed by McLaughlin and Scott provided evolution equations for the collective parameters of a perturbed sine-Gordon fluxon \cite{scott}. Ishimori and Munakata applied this approach to the discrete sine-Gordon chain and showed that the PN potential produces kink wobbling or pinning; they also found a first-order dressing that steepens the kink and an asymmetric radiative correction, with stronger backward than forward radiation \cite{ishimori}. In the strongly discrete regime, Peyrard and Kruskal found numerically that moving kinks radiate, lose memory of their initial velocity, and approach preferred propagation velocities; they also identified mobile multiple-kink excitations in regimes where ordinary kinks are pinned \cite{kruskal}. A lattice Hamiltonian formulation was developed by Willis, El-Batanouny, and Stancioff \cite{willis}, while Boesch, Willis, and El-Batanouny explicitly analysed the spontaneous emission of radiation by a discrete sine-Gordon kink \cite{willis2}. These studies separate two kinds of discreteness effects: regular corrections to the smooth kink profile and nonperturbative phenomena such as pinning and resonant radiation. The present work concentrates on the former.

An important complementary direction is the construction of exceptional discretisations that suppress the PN barrier. Speight constructed topological discrete Klein--Gordon models that preserve the continuum lower energy bound and admit static kinks centred arbitrarily relative to the lattice \cite{speight2}. Balmforth, Craster, and Kevrekidis introduced a discrete Evans function for determining the linear spectrum, stability, and internal modes of localised lattice states \cite{kevrekidis4}. In the dipole-chain realisation studied by Speight and Zolotaryuk, the absence of the PN barrier makes single-kink dynamics unusually continuum-like, although exact finite-speed travelling kinks were not found except possibly at very low velocity~\cite{speight}. Oxtoby, Pelinovsky, and Barashenkov subsequently showed by beyond-all-orders analysis that translationally invariant static kinks do not generally imply radiationless motion: sliding kinks occur only at isolated velocities in the exceptional $\phi^4$ discretisations they examined \cite{oxtoby}. Roy, Dmitriev, Kevrekidis, and Saxena compared five $\phi^4$ discretisations through their kink profiles, linear spectra, and kink--antikink collisions, finding model-dependent broadening or narrowing and an overall increase in collision elasticity away from the continuum limit \cite{kevrekidis2}. Dmitriev and collaborators also constructed generalised discrete $\phi^4$ models with exact, stable tanh-type kinks moving at prescribed, arbitrarily high velocities \cite{kevrekidis1}. Together, these results show that both kink shape and mobility depend sensitively on the chosen discretisation.

Related localised waves arise in nonlinear electrical transmission lines. Numerical finite-difference studies of series-connected discrete Josephson transmission lines demonstrated cutoff propagation, dispersion, and shock-wave formation \cite{mohebbi}. Coupled-envelope descriptions produced solitary-wave and kink signals in discrete nonlinear dispersive lines~\cite{malomed}, while exact travelling kink and antikink solutions were derived for nonlinear electrical lattices and verified by direct simulation \cite{abdoul}. For integrable self-dual transmission-line equations, Bogdan and Laptev constructed explicit kink--breather and breather--breather solutions and calculated collision-induced position and phase shifts \cite{bogdan}. Dutykh and Caputo formulated conservation-based coupling conditions and a symplectic scheme for sine-Gordon kinks and breathers propagating on networks \cite{denys}. Solitons in quantum Josephson transmission lines have also been proposed as moving effective horizons for analogue Hawking radiation \cite{katayama}.

For the series-connected JTL considered below, earlier work derived shock profiles and analysed the effects of resistive elements \cite{kogan1}. A quasi-continuum reduction was then used to obtain supersonic kinks and solitons, their velocities and profiles, Korteweg--de Vries and modified Korteweg--de Vries limits for weak waves, and stationary shocks in dissipative lines \cite{kogan2}. Subsequent comparison of continuum, quasi-continuum, and quasi-discrete descriptions clarified that the compact travelling waves of a lossless line are kinks and solitons and introduced a simple-wave approximation for right- and left-going waves and shock formation \cite{kogan3}. The present calculation extends this line of work by retaining the next term in the long-wavelength expansion and determining its correction to the kink profile.

The same long-wavelength philosophy is widely used for Fermi--Pasta--Ulam--Tsingou-type lattices. Vainchtein reviewed the existence and structure of solitary waves in FPU-type chains \cite{anna}. More recently, Yang, Sun, Liu, and Kevrekidis derived a modified KdV reduction supporting solitons, breathers, and rational localised waves on homogeneous and periodic backgrounds, and verified these quasi-continuum patterns against direct FPUT simulations \cite{kevrekidis3}. These results further motivate a systematic assessment of higher-order continuum corrections in discrete nonlinear systems.

\section{The Kinks in Nonlinear Klein--Gordon Chains}
\label{kk}

Consider nonlinear Klein--Gordon (KG) chains described by
\begin{eqnarray}
\label{pop}
\frac{d^2 \phi_n}{d \tau^2}-\frac{1}{h^2}D_2(\phi_n)=F\left(\phi_n\right),
\end{eqnarray}
where $D_2$ is the discrete second derivative
\begin{eqnarray}
D_2(\phi_n):= \phi_{n+1}-2\phi_{n}+\phi_{n-1},
\end{eqnarray}
\indent Here
 $F(\phi)$ is a nonlinear dimensionless function, $\tau$ is dimensionless time measured in units of the inverse frequency $\omega_0$ of small oscillations about equilibrium (possibly up to a numerical factor of order unity), and $h$ is the dimensionless lattice period measured in units of $c_0/\omega_0$, where $c_0$ is the velocity of smooth, small-amplitude (harmonic) waves.

To pass from difference equations to differential equations, we replace the index $n$ by the continuous variable $z=nh$ and approximate the discrete second derivative in Equation~(\ref{pop}) by an infinite Taylor series of partial derivatives,
\begin{eqnarray}
\label{comm33}
D_2(\phi_n)=
h^2\frac{\partial^2\phi}{\partial z^2}+\frac{h^4}{12}\frac{\partial^4\phi}{\partial z^4}+\frac{h^6}{360}\frac{\partial^6\phi}{\partial z^6}+\dots
\end{eqnarray}
\indent Thus Equation (\ref{pop}) becomes the equation in partial derivatives
\begin{eqnarray}
\label{pp}
\frac{\partial^2 \phi}{\partial\tau^2}-\frac{\partial^2\phi}{\partial z^2}-\frac{h^2}{12}\frac{\partial^4\phi}{\partial z^4}-\dots=F\left(\phi_n\right).
\end{eqnarray}
\indent Equation (\ref{pp}) follows from the Lagrangian
\begin{eqnarray}
L&=&\int \Bigg[\frac{1}{2}\left(\frac{\partial\phi}{\partial \tau}\right)^2
-\frac{1}{2}\left(\frac{\partial\phi}{\partial z}\right)^2
+\frac{h^2}{24}\left(\frac{\partial^2\phi}{\partial z^2}\right)^2\nonumber\\
&-&\frac{h^4}{720}\left(\frac{\partial^3\phi}{\partial z^3}\right)^2
+\dots-\Pi(\phi)\Bigg]dz,
\end{eqnarray}
where
\begin{eqnarray}
\label{pb77}
\Pi(\phi)=-\int  F(\phi) d\phi.
\end{eqnarray}

We seek travelling-wave solutions of the form
\begin{eqnarray}
\label{rund}
\phi(z,\tau)=\phi(x),
\end{eqnarray}
where $\theta=U\tau-z$, and $U$ is the wave velocity. For these waves,  partial derivatives  in (\ref{pp}) become the ordinary derivatives and $d/d\tau$  turns into $Ud/dz$.

In what follows, $\alpha^2$ is treated as a small parameter.
At leading order (LO) with respect to this parameter, we truncate the series (\ref{comm33}) after its first term. At next-to-leading order (NLO), we retain the second term as well \cite{willis}, and Equation (\ref{pp}) becomes
\begin{eqnarray}
\label{mm33}
\frac{d^2 \phi}{d\theta^2}+\frac{\alpha^2}{2}\frac{d^4 \phi}{d\theta^4}=-F\left(\phi\right),
\end{eqnarray}
where
\begin{subequations}
\begin{alignat}{4}
\alpha^2&:= \frac{h^2}{6(1-U^2)^2}\\
\theta&= \frac{x}{\sqrt{1-U^2}}.
\end{alignat}
\end{subequations}

Multiplying Equation (\ref{mm33}) by $d\phi/d\theta$ and integrating gives
\begin{eqnarray}
\label{pb}
\frac{1}{2}\left\{1+\alpha^2pS[\phi(\theta)]\right\}\left(\frac{d\phi}{d\theta}\right)^2=\Pi(\phi),
\end{eqnarray}
where  $pS$ is the pseudo-Schwarzian derivative (PSD)
\begin{eqnarray}
pS[\phi(\theta)]:=\frac{d^3\phi/d\theta^3}{d\phi/d\theta}
-\frac{1}{2}\left(\frac{d^2\phi/d\theta^2}{d\phi/d\theta}\right)^2;
\end{eqnarray}
\indent Some identities involving this derivative are presented in Appendix \ref{weak2}.

We consider kinks satisfying the boundary conditions
\begin{eqnarray}
\label{grand}
\lim_{\theta\to -\infty}\phi(\theta)=\phi_2\;,\hskip 1cm
\lim_{\theta\to +\infty}\phi(\theta)=\phi_1\,.
\end{eqnarray}
where $\phi_1$ and $\phi_2$ are two zeros of $F(\phi)$
\begin{eqnarray}
F(\phi_1)=F(\phi_2)=0.
\end{eqnarray}
\indent Equation (\ref{pb}) implies that kinks exist only if $\Pi(\phi_1)=\Pi(\phi_2)$. The indefinite integral in Equation (\ref{pb77}) should therefore be interpreted as a definite integral with limits $\phi_1$ and $\phi$.

 In the LO  approximation,  Equation (\ref{pb}) takes the form
\begin{eqnarray}
\label{pb2}
\frac{1}{2}\left(\frac{d\phi}{d\theta}\right)^2 =\Pi(\phi).
\end{eqnarray}
\indent At NLO, the PSD on the left-hand side of Equation (\ref{pb}) must be included, but it may be evaluated using the solution of Equation (\ref{pb2}). We now consider two important special cases. These models are of interest in their own right and are also relevant to the second part of the paper, which concerns a series-connected Josephson transmission line.

\subsection{$\phi^4$ Chain}

For the  $\phi^4$ chain
\begin{eqnarray}
\label{pbb}
\Pi(\phi)=\frac{1}{2}\left(1-\phi^2\right)^2
\end{eqnarray}
and  the solution of (\ref{pb2}) is
\begin{eqnarray}
\label{tt}
\phi=\tanh\theta
\end{eqnarray}
(we have chosen $\phi_1=-\phi_2=1$).
Equations (\ref{pb6})  and  (\ref{pb7}) (for $n=2$) in this case are
\begin{eqnarray}
\label{19}
\left(\frac{d\phi}{d\widetilde{\theta}}\right)^2 =\left(1-\phi^2\right)^2,
\end{eqnarray}
where
\begin{subequations}
\begin{alignat}{4}
\widetilde{\theta}&=\frac{\sinh^{-1}\left[m\sinh\theta\right]}{\sqrt{1+2\alpha^2}}\\
m^2&=\frac{1+2\alpha^2}
{1-2\alpha^2}.
\end{alignat}
\end{subequations}
\indent The solution of (\ref{19})    is
\begin{eqnarray}
\label{v1b}
\phi(x)=\tanh \widetilde{\theta}.
\end{eqnarray}

\begin{figure}[tbp]
\includegraphics[width=\columnwidth]{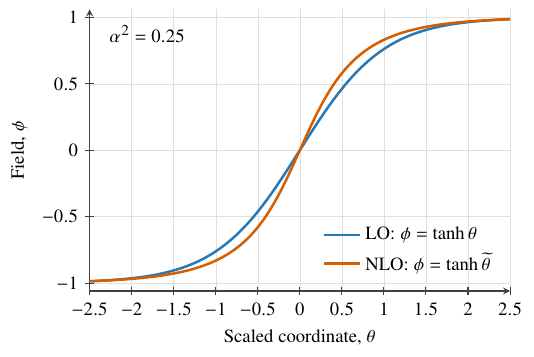}
\caption{LO
 and NLO kink profiles in the $\phi^4$ chain. The curves correspond to \mbox{Equations (\ref{tt}) and (\ref{v1b}),} respectively, for $\alpha^2=0.25$.}
 \label{fi4}
\end{figure}

Figure \ref{fi4} compares the NLO kink profile for the $\phi^4$ chain [Equation (\ref{v1b})] at $\alpha^2=0.25$ with the LO profile [Equation (\ref{tt})]. We see that the  term with the fourth-order derivative in Equation (\ref{mm33}) compresses the central part of the kink while broadening its tails.

\subsection{Sine-Gordon Chain}

For the sine-Gordon chain
\begin{eqnarray}
\Pi(\phi)=1-\cos\phi,
\end{eqnarray}
and for the solution of (\ref{pb2}) we obtain
\begin{subequations}
\begin{alignat}{4}
\phi&=4\tan^{-1}\left(\exp\theta\right) \label{ttt}\\
\frac{d\phi}{d\theta}&=\frac{2}{\cosh\theta}
\end{alignat}
\end{subequations}
(we have chosen $\phi_2=0$, $\phi_1=2\pi$).
Equations (\ref{pb6})  and  (\ref{pb7}) (for $n=1$) in this case are
\begin{eqnarray}
\label{pb9}
\left(\frac{d\phi}{d\widetilde{\theta}}\right)^2=4\sin^2(\phi/2),
\end{eqnarray}
where
\begin{subequations}
\begin{alignat}{4}
\widetilde{\theta}&=\frac{\sinh^{-1}\left[m\sinh\theta\right]}{\sqrt{1+\alpha^2/2}}\\
m^2&=\frac{1+\alpha^2/2}{1-\alpha^2}.
\end{alignat}
\end{subequations}
\indent The solution of (\ref{pb9})   is
\begin{eqnarray}
\label{k}
\phi(x)=4\tan^{-1}\left(\exp\widetilde{\theta}\right).
\end{eqnarray}

To compare Equation (\ref{k}) with the result of Ref. \cite{ishimori}, we expand it in $\alpha$ and retain the first two terms,
\begin{eqnarray}
\label{kkk}
\phi=4\tan^{-1}\left(\exp\theta\right)+\frac{\alpha^2}{2}\left(3\tanh\theta-\theta\right)\sech\theta.
\end{eqnarray}
\indent Equation (\ref{kkk}) exactly coincides with  Equation (4.5) of Ref. \cite{ishimori}.
The Peierls--Nabarro potential barrier, wobbling, and kink deceleration \cite{scott,ishimori,willis2} are nonperturbative effects and are therefore absent from the present perturbation theory.

\begin{figure}[tbp]
\includegraphics[width=\columnwidth]{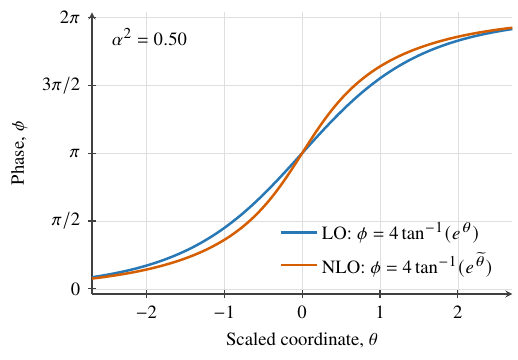}
\caption{LO and NLO kink profiles in the sine-Gordon chain. The curves correspond to \mbox{Equations (\ref{ttt}) and (\ref{k}),} respectively, for $\alpha^2=0.5$.}
 \label{sg}
\end{figure}

Figure \ref{sg} compares the NLO kink profile for the sine-Gordon chain [Equation (\ref{k})] at $\alpha^2=0.5$ with the LO profile [Equation (\ref{ttt})]. We see that the  term with the fourth-order derivative in Equation (\ref{mm33}) compresses the central part of the kink while broadening its tails.

\section{Kinks in a Discrete Series-Connected Josephson Transmission Line}
\label{weak}

Circuits containing Josephson junctions (JJs) \cite{josephson}, known as Josephson transmission lines (JTLs) \cite{barone,pedersen}, provide interesting and potentially important examples of nonlinear transmission lines. Kinks propagating in series-connected JTLs have previously been studied in the LO approximation \cite{katayama,kogan2}. In this section, we treat such kinks in the NLO approximation \cite{kogan3}.

\begin{figure}[tbp]
\includegraphics[width=\columnwidth]{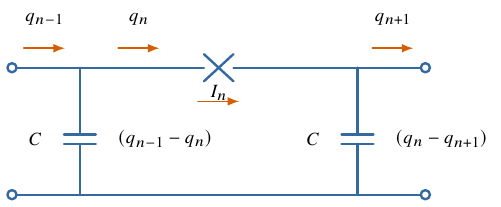}
\caption{Unit cell of the discrete series-connected JTL.}
\label{trans1}
\end{figure}

Consider the JTL model shown in Figure \ref{trans1}, constructed from identical JJs and capacitors.

We choose the Josephson phases $\varphi_n$ and the charges $q_n$ that have passed through the JJs as dynamical variables. The circuit equations are
\begin{subequations}
\label{ave7}
\begin{alignat}{4}
\frac{\hbar}{2e}\frac{d \varphi_n}{d\tau}&=\frac{1}{C}\left(q_{n+1}-2q_{n}+q_{n-1}\right) ,\label{ave7a}\\
\frac{dq_n}{d\tau} &= I_n ,\label{ave7b}
\end{alignat}
\end{subequations}

where $C$ is the capacitance,  and
\begin{eqnarray}
\label{jo}
I= I(\varphi)
\end{eqnarray}
where $I(\varphi)$ is an odd function.

Measuring distance in units of the JTL period, current through a JJ in units of \mbox{$I_c=I'(\varphi)|_{\varphi=0}$,} time in units of $\sqrt{L_JC}$, and charge in units of $I_c\sqrt{L_JC}$, where $L_J=\hbar/(2eI_c)$, renders Equation (\ref{ave7}) dimensionless:
\begin{subequations}
\label{7}
\begin{alignat}{4}
\frac{d \varphi_n}{d\tau}&=D_2(q_n) ,\label{7a}\\
\frac{dq_n}{d\tau} &= I_n.
\end{alignat}
\end{subequations}

As in Section \ref{kk}, we replace the index $n$ by a continuous variable $z$ and approximate the discrete second derivative by the infinite Taylor series (\ref{comm33}). Equation (\ref{7}) then becomes
\begin{subequations}
\label{a7}
\begin{alignat}{4}
\frac{\partial \varphi}{\partial \tau}&=\frac{\partial^2q}{\partial z^2}+\frac{1}{12}\frac{\partial^4q}{\partial z^4}+\frac{1}{360}\frac{\partial^6q}{\partial z^6}+\dots \label{a7a}\\
\frac{\partial q}{\partial \tau} &=I(\varphi)  \label{a7b}
\end{alignat}
\end{subequations}
\indent The expansion parameter is identified below.

Equation (\ref{a7}) can be obtained from the Hamiltonian \cite{kogan1}
\begin{eqnarray}
\label{hamilton}
H=\int \left[\int  I(\varphi) d\varphi+\frac{1}{2}\left(\frac{\partial q}{\partial z}\right)^2
\right.\nonumber\\
\left.-\frac{1}{24}\left(\frac{\partial^2q}{\partial z^2}\right)^2
+\frac{1}{720}\left(\frac{\partial^2q}{\partial z^2}\right)^2-\dots\right]dz,
\end{eqnarray}
$\varphi$ and $q$ being conjugate variables.
 Unlike the case of the KG chains, the LO approximation for the JTL retains the first two terms of the series (\ref{comm33}) (the first three terms in the \mbox{Hamiltonian (\ref{hamilton}))}, whereas the NLO approximation keeps additionally the third
term of the series (\ref{comm33}) (the  fourth terms in the Hamiltonian (\ref{hamilton})). The JTL NLO approximation has an advantage over both the corresponding LO approximation and the NLO approximation for the KG chains. To see this, write the potential part of the Hamiltonian in Fourier space as
\begin{eqnarray}
\label{integra}
E_p=\int \left(\frac{\omega^2}{2}-\frac{\omega^4}{24}+\frac{\omega^6}{720}-\dots\right)
q^2(\omega)d\omega.
\end{eqnarray}
\indent When truncated after the third term, the integrand in (\ref{integra})  is strictly positive; this is not the case when it is truncated after the second term \cite{kogan2,kogan3}.

Smooth, small-amplitude harmonic waves on the constant background $\varphi_1$, described by Equation (\ref{a7}) with Equation (\ref{a7a}) truncated after the first term, obey the linear dispersion relation
$\omega=u(\varphi_1)k$,
where $u$ is both the phase and group velocity of these waves,
\begin{eqnarray}
\label{disper}
u^2(\varphi_1)=I^{\prime}(\varphi_1).
\end{eqnarray}
\indent In the most important practical case, the Josephson relation is
\begin{eqnarray}
\label{sin}
I=\sin\varphi,
\end{eqnarray}
where $I_c$ is the JJ critical current, and
\begin{eqnarray}
u^2(\varphi)&=\cos\varphi.
\end{eqnarray}

When studying the kinks, it is more convenient to use $I$, rather than $\varphi$, as a dynamical variable and to invert the Josephson relation (\ref{jo})
 \cite{mohebbi,katayama,kogan2,kogan3}:
\begin{eqnarray}
\varphi=\varphi(I).
\end{eqnarray}
\indent In general, $\varphi$ is a multi-valued function of $I$. Here, however, we consider weak kinks for which the current varies only near zero, so a single branch suffices.

General travelling-wave solutions have the form
\begin{eqnarray}
\label{run}
I(z,\tau)=I(x),\hskip 1cm q(z,\tau)=q(x),
\end{eqnarray}
where $x=U\tau-z$, and $U$ is the wave velocity.
(The dimensionless velocities may be converted to physical velocities using the units of distance and time defined after \mbox{Equation (\ref{jo})}.) For these waves, partial derivatives reduce to ordinary derivatives, and Equation (\ref{a7}) becomes the following ordinary differential equation for the current:
\begin{eqnarray}
\label{v7}
U^2\frac{d\varphi}{d x}=\frac{dI}{d x}+\frac{1}{12}\frac{d^3I}{d x^3}+\frac{1}{360}\frac{d^5I}{d x^5}.
\end{eqnarray}
\indent Integration with respect to $x$ gives
\begin{eqnarray}
\label{v99}
\frac{1}{12}\frac{d^2I}{d x^2}+\frac{1}{360}\frac{d^4I}{d x^4}
=U^2\varphi(I)-I+F,
\end{eqnarray}
where $F$ is the constant of integration.
Multiplying Equation (\ref{v99}) by $dI/dx$ and integrating~gives
\begin{eqnarray}
\label{p99b}
\frac{1}{2}\left[1+\frac{1}{15}pS[I(x)]\right]\left(\frac{dI}{d x}\right)^2 =\Pi(I),
\end{eqnarray}
where
\begin{eqnarray}
\label{p9}
\Pi(I)=12\int\left[U^2\varphi(I)-I+F\right]d I
\end{eqnarray}

We are interested in the travelling  waves defined by
the boundary conditions
\begin{eqnarray}
\label{gran}
\lim_{x\to -\infty}I(x)=I_2\;,\hskip 1cm
\lim_{x\to +\infty}I(x)=I_1\,.
\end{eqnarray}
\indent Under these boundary conditions, Equation (\ref{p99b}) has two types of solutions \cite{kogan2}: kinks, for which $I_2=-I_1$, and solitons, for which $I_2=I_1$. (This statement, like Equation (\ref{ss}) below, remains valid even if we
keep in the r.h.s. of (\ref{a7a}), and hence in the r.h.s. of  (\ref{v7}), the whole infinite series \cite{kogan3}.)
In both cases, the indefinite integral in Equation (\ref{p9}) should be interpreted as a definite integral with limits $I_1$ and $I$. Solitons are considered in Appendix~\ref{weak3}.

For kinks, the boundary conditions determine the constants entering Equation (\ref{p99b}):
\begin{subequations}
\label{ss}
\begin{alignat}{4}
F&=0 \label{felo}\\
U^2(I_1)&=I_1/\varphi(I_1). \label{velo}
\end{alignat}
\end{subequations}

Using Equation (\ref{ss}), we can write Equation (\ref{p9}) as
\begin{eqnarray}
\label{p10}
\Pi(I)=12\int_{I_1}^I\left[\frac{\varphi(I)/I}{\varphi(I_1)/I_1}-1\right]Id I
\end{eqnarray}

Substituting the expansion of $\varphi(I)$ in powers of the current into the integrand and evaluating the integral gives
\begin{eqnarray}
\label{expo}
\Pi(I)=\frac{v^2}{2}\left(I^2-I_1^2\right)^2\nonumber\\
\left[1 +\frac{2wv^2-5v^2}{30}I_1^2+\frac{wv^2}{30}I^2+\dots\right],
\end{eqnarray}
where the ellipsis denotes terms of fourth and higher order, and
\begin{subequations}
\begin{alignat}{4}
v^2:&=\varphi^{(III)}(0)\\
w:&=\frac{\varphi^{(V)}(0)}{\left[\varphi^{(III)}(0)\right]^2}.
\end{alignat}
\end{subequations}
\indent Note that expanding (\ref{disper}) we obtain
\begin{eqnarray}
\label{speco}
1/u^2(I)=1+\frac{1}{2}\varphi^{(III)}(0)I_1^2,
\end{eqnarray}
and expanding (\ref{velo}) we obtain
\begin{eqnarray}
1/U^2(I_1)&=1+\frac{1}{6}\varphi^{(III)}(0)I_1^2.
\end{eqnarray}
\indent Only supersonic kinks exist, $U(I_1)>u(I_1)$ \cite{kogan2}, which implies that $\varphi^{(III)}(0)>0$. (The dispersion relation for smooth, small-amplitude waves
on the stable equilibrium background is $\omega^2=k^2+4$ in the $\phi^4$  chain, and $\omega^2=k^2+1$ in the sine-Gordon chain.  In both cases the phase velocity $d\omega/dk$ at $k=0$ is zero. So the kinks in these systems moving at any nonzero velocity are supersonic.)

\subsection*{Weak Kinks}
\label{weak4}

For the KG chains, the kink amplitude is fixed and the kink velocity is a free parameter. For the JTL, Equation (\ref{velo}) shows that the kink velocity and amplitude are related. We will use the  kink amplitude as the free parameter, which is assumed to be small.

In the LO approximation
\begin{eqnarray}
\label{expon}
\Pi(I)=\frac{v^2}{2}\left(I^2-I_1^2\right)^2.
\end{eqnarray}
\indent Comparing Equations (\ref{pb}) and (\ref{pbb}) with Equations (\ref{p99b}) and (\ref{expon})
we see that after introducing
\begin{eqnarray}
\label{te}
\theta =vI_1x,
\end{eqnarray}
the equations for the JTL become nearly identical to those for the $\phi^4$ chain, provided that we make the identifications
\begin{subequations}
\label{iden}
\begin{alignat}{4}
I/I_1&\leftrightarrow\phi\\
\frac{v^2}{15}I_1^2&\leftrightarrow \alpha^2.
\end{alignat}
\end{subequations}
\indent Thus the expansion parameter  in Equation (\ref{a7a}) (same as in Equation (\ref{expo})) is $vI_1$. The physical interpretation of the formal requirement $vI_1\ll1$: over the relevant current range, $\varphi(I)$ must deviate only weakly from linearity.
Equation (\ref{te}) also shows that the perturbation theory expansion parameter in the present case can equivalently be interpreted, as for the KG chains, as the ratio of the JTL period to the kink width.

In distinction from (\ref{19}), the r.h.s. of Equation (\ref{p99b}) contains the higher-order correction. This correction can be approximated by substituting the LO solution into the brackets in Equation (\ref{expo}). In the present case, the LO solution is
\begin{eqnarray}
\label{20}
I(x)=I_1\tanh \theta.
\end{eqnarray}
\indent Equation (\ref{p99b}) then becomes
\begin{eqnarray}
\label{189b}
\left[1+2\alpha^2(1-2\sech^2\theta)\right]I_1^2
\left(\frac{dI}{d \theta}\right)^2=\nonumber\\
 =\left(I^2-I_1^2\right)^2 \left[1 +\frac{\alpha^2}{2}\left(2w-5+w\sech^2\theta\right)\right].
\end{eqnarray}
\indent The solution of (\ref{189b}) in the NLO approximation is
\begin{eqnarray}
\label{v12}
I(x)=I_1\tanh\widetilde{\theta},
\end{eqnarray}
where
\begin{eqnarray}
\label{v9}
\widetilde{\theta}:=
\int\left[1+\frac{\alpha^2}{4}\left(3w-9+\frac{8-w}{\cosh^2\theta}\right)\right]d\theta\nonumber\\
=\left[1+\frac{\alpha^2}{4}\left(3w-9\right)\right]\theta+\frac{\alpha^2}{4}(8-w)\tanh \theta.
\end{eqnarray}

For the canonical Josephson relation (\ref{sin})
\begin{subequations}
\begin{alignat}{4}
v^2&=1\\
w&=9
\end{alignat}
\end{subequations}
and Equation (\ref{v12}) becomes
\begin{eqnarray}
\label{v14}
I(x)=I_1\tanh\left[I_1\left(1+\frac{3I_1^2}{10}\right)x-\frac{I_1^3}{60}\tanh\left(I_1x\right)\right].
\end{eqnarray}
\indent Thus, the term with the sixth-order derivative in the r.h.s. of Equation (\ref{a7a})
 compresses the~kink.

\section{Conclusions}

We have developed a next-to-leading-order (NLO) long-wavelength description of travelling kinks in three representative nonlinear lattices: the $\phi^4$ and sine-Gordon Klein--Gordon (KG) chains and a series-connected discrete Josephson transmission line (JTL). In each case, the discrete spatial operator was expanded one order beyond the conventional continuum approximation. For the KG chains this produces a fourth-order spatial correction, whereas for the JTL the consistent Hamiltonian truncation retains the sixth-order term as well. The resulting travelling-wave equations can be integrated in a common form involving the pseudo-Schwarzian derivative. This structure provides a unified perturbative method for calculating profile corrections in otherwise rather different lattice models.

For the KG chains, the method yields closed-form NLO profiles obtained by replacing the leading-order travelling coordinate by a nonlinear, profile-dependent coordinate. In the $\phi^4$ model, the correction compresses the central part of the kink while broadening its wings; in the sine-Gordon model, it compresses the central part of the $2\pi$ phase kink. Expanding the sine-Gordon result reproduces the previously obtained first-order correction~\cite{ishimori}, providing a check on the calculation. The analysis therefore shows explicitly that the first regular effect of discreteness is not merely a change in the kink velocity or a uniform rescaling of its width: different parts of the profile are deformed differently, and the deformation depends on the nonlinear on-site potential.

For the series-connected JTL, the kink amplitude and velocity are not independent. The boundary conditions determine the velocity through Equation (\ref{velo}), while the small kink amplitude may be used as the perturbation parameter. In the weak-kink regime, the JTL equation maps closely onto the $\phi^4$ calculation, and the NLO correction again compresses the kink. For the canonical Josephson current--phase relation, this correction is given explicitly by Equation (\ref{v14}). Equations (\ref{v9}) and (\ref{v14}) also show that the apparently different expansion parameters used for the KG chains and the JTL have the same physical meaning: they measure the ratio of the lattice period to the characteristic kink width.

The corrections found here remain comparatively modest even when the nonlinearity is appreciable, which supports the usefulness of the quasi-continuum approximation over a nontrivial parameter range. Its validity is nevertheless controlled by kink width rather than by nonlinearity alone. As the kink approaches the lattice scale, the higher-order expansion ceases to be ordered; the singularity of the NLO formulas at the bounds discussed in Appendix \ref{weak2} is an internal indication of this breakdown. Moreover, the present power-series treatment describes regular profile corrections but does not capture effects that are beyond all algebraic orders in the lattice spacing, including the Peierls--Nabarro barrier, pinning, and resonant radiation \cite{ishimori,kruskal,willis2,oxtoby}.

The framework can be extended in several directions. Direct comparison with travelling waves of the original difference equations would quantify the NLO accuracy and the onset of lattice-scale effects. Linear-stability and radiation calculations could determine how the corrected profiles influence mobility and energy loss, while dissipation and circuit nonuniformity could be incorporated for experimentally realistic JTLs. It would also be useful to apply the same pseudo-Schwarzian formulation to other Klein--Gordon, FPUT, and nonlinear transmission-line lattices. Such extensions would clarify which profile corrections are universal consequences of discreteness and which depend on the specific lattice Hamiltonian and nonlinear force law.

\appendix
\section{The Pseudo-Schwarzian Derivative}
\label{weak2}

If
\begin{eqnarray}
\label{b1}
\frac{d\phi}{d\theta}=\frac{1}{\cosh^n\theta},
\end{eqnarray}
then
\begin{subequations}
\label{b2}
\begin{alignat}{4}
\frac{d^2\phi}{d\theta^2}&=-\frac{n\tanh \theta}{\cosh^n\theta}\\
\frac{d^3\phi}{d\theta^3}&=\frac{n^2}{\cosh^n\theta}
-\frac{n(n+1)}{\cosh^{n+2}\theta}.\label{b2b}
\end{alignat}
\end{subequations}
\indent Hence for the function satisfying Equation (\ref{b1})
\begin{eqnarray}
\label{pu4}
pS[\phi(\theta)]=\frac{1}{2}\left[n^2-n(n+2)\sech^2\theta\right].
\end{eqnarray}
\indent Thus Equation (\ref{pb}) both for the $\phi^4$ and for the sine-Gordon chains  we can approximate as
\begin{eqnarray}
\label{pb66}
\frac{1}{2}\left[1+\frac{n^2\alpha^2}{2}-\frac{n(n+2)\alpha^2}{2}\sech^2\theta\right]
\left(\frac{d\phi}{d \theta}\right)^2&=\Pi(\phi),
\end{eqnarray}
where $n=2$ in the former case and $n=1$ in the latter.
A good perturbation theory should contain within itself the indication to how it can break down. Ours satisfies this criterium---Equation (\ref{pb66}) breaks down for $\alpha^2>1/n$,
which corresponds to $U^2>1-h\sqrt{n/6}$ for the KG chains.

Equation (\ref{pb66})  can be written down as
\begin{eqnarray}
\label{pb6}
\frac{1}{2}\left(\frac{d\phi}{d \widetilde{\theta}}\right)^2=\Pi(\phi),
\end{eqnarray}
where
\begin{subequations}
\label{pb7}
\begin{alignat}{4}
\widetilde{\theta}&=\frac{\sinh^{-1}\left[m(\alpha)\sinh\theta\right]}{\sqrt{1+n^2\alpha^2/2}}
\label{pb7a}\\
m^2(\alpha):&=\frac{1+n^2\alpha^2/2}{1-n\alpha^2}.
\end{alignat}
\end{subequations}

Expanding (\ref{pb7a}) with respect to $\alpha$ we obtain
\begin{eqnarray}
\label{1278}
\widetilde{\theta}=\theta
+\frac{n\alpha^2}{4}\left[(n+2)\tanh\theta-n\theta\right]+\dots.
\end{eqnarray}
\indent From Equation (\ref{1278}) it follows that the higher-order terms compress the central (where $\tanh \theta$ can be approximated as $\theta$) part of the kinks in the considered KG chains but make their wings (where $\tanh \theta$ can be ignored) broader.

Also using Equation  (\ref{pu4}) we can check up that
(\ref{tt}) is the exact solution \cite{triki,susanto} of  Equation (\ref{pb}) (though not of the original problem) for
\begin{eqnarray}
\Pi(\phi)=\frac{1}{2}\left(1-\phi^2\right)^2\left(1-4\alpha^2+8\alpha^2\phi^2\right),
\end{eqnarray}
and  (\ref{ttt})---for
\begin{eqnarray}
\Pi(\phi)=\sin^2\frac{\phi}{2}\left(2+\alpha^2-3\alpha^2\sin^2\frac{\phi}{2}\right).
\end{eqnarray}

\section{Weak Solitons in a Discrete JTL}
\label{weak3}

Let us consider  weak solitons ($|I-I_1|\ll |I_1|\sim 1$) (so multi-valuedness of $\varphi(I)$ is not a problem). The calculations in this Appendix will be performed in the LO approximation (in distinction from the rest of the paper).
So we discard the fourth-order derivative  in the l.h.s. of Equation (\ref{v99}), expand the r.h.s. of the equation
with respect to the powers of $\tilde{I}=I-I_1$ and truncate the series after the square term, thus approximating  (\ref{v99}) as
\begin{eqnarray}
\label{v999}
\frac{d^2\widetilde{I}}{d x^2}=\beta^2\widetilde{I}+\frac{\gamma}{2}\widetilde{I}^2,
\end{eqnarray}
where
\begin{subequations}
\begin{alignat}{4}
\beta^2:&=\frac{U^2}{u^2(I_1)}-1\\
\gamma:&=U^2\varphi^{\prime\prime}(I_1)
\end{alignat}
\end{subequations}
(only supersonic solitons exist \cite{kogan2}).
Using Equations (\ref{b1}) and (\ref{b2b}) for $n=2$ we can check up
that  soliton \cite{kogan2}
\begin{eqnarray}
\label{solist}
\widetilde{I}=-\frac{3\beta^2}{\gamma\cosh^2\left(\beta x/2\right)}
\end{eqnarray}
is the solution of Equation (\ref{v999}). The soliton is dark if $I_1\varphi^{\prime\prime}(I_1)>0$, and bright if  \mbox{$I_1\varphi^{\prime\prime}(I_1)<0$.}
For the Josephson law (\ref{sin})
\begin{eqnarray}
\varphi^{\prime\prime}(I_1)=\frac{I_1}{(1-I^2_1)^{3/2}}
\end{eqnarray}
and the soliton is dark.

In this  section, similar to what was done for the KG chains, we considered the velocity of the soliton as a free parameter; the amplitude of the soliton is connected with the velocity
by Equation (\ref{solist}).
So the  solitons are weak if they move slightly faster than small and smooth waves on the same constant background.

The discarded terms  in the r.h.s. of Equation (\ref{v999})
contained higher powers of  $\beta^2$ (than those we kept). The same can be said about  the discarded terms in the expansion (\ref{comm33}).
Looking at Equation (\ref{solist}) we understand that the expansion parameter in this section
 can be understood in the same sense as in the rest of the paper---as the ratio of the chain period to the soliton width.

\end{document}